\documentstyle[preprint,aps]{revtex}
\newcommand{\bc}{\begin{center}}
\newcommand{\ec}{\end{center}}
\newcommand{\bfig}{\begin{figure}}
\newcommand{\efig}{\end{figure}}
\newcommand{\beq}{\begin{equation}}
\newcommand{\eeq}{\end{equation}}
\newcommand{\beqa}{\begin{eqnarray}}
\newcommand{\eeqa}{\end{eqnarray}}

\begin{document}
\draft
\title{
Nesting Mechanism for d-symmetry Superconductors\footnote{
Research supported by DOE grant No.DE-FG05-84ER45113 and by HNRF grants
OTKA 2950 and T4473}}
\author{
J. Ruvalds, C.T. Rieck\footnote{Present address :
Abteilung f\"{u}r Theoretische Festk\"{o}rperphysik, Universit\"{a}t Hamburg,
Hamburg, Germany.},
S. Tewari, J. Thoma}
\address{Physics Department,
University of Virginia,
Charlottesville, VA 22901}
\author{
A. Virosztek}
\address{Research Institute for Solid State Physics,
1525 Budapest 114, POB 49, Hungary}

\maketitle
\begin{abstract}
A nested Fermi surface with nearly parallel orbit segments is found to yield
a singlet
d--wave  superconducting state at high temperatures for a restricted range
of the on--site Coulomb repulsion that
avoids the competing spin density wave instability. The computed
superconducting transition temperature drops dramatically
as the nesting vector is decreased, in accord with recent photo--emission
data on the Bi2212 and Bi2201 cuprates.
\end{abstract}

\newpage

	Even though the BCS theory \cite{bcs} provides a successful description of
conventional superconductors,
the concept of alternate electron pairing states of finite angular momentum
has evolved \cite{pwa}
to encompass various physical systems. Spin fluctuations
suppress \cite{don,berk} the BCS isotropic pair binding induced by phonon
exchange, and hence materials with strong
repulsive interactions are prospects for anisotropic
pairing. Superfluid He$^3$ exhibits p-wave pairing \cite{legg}, and heavy
Fermion superconductors offer another unconventional case
at very low temperatures.

	Copper oxides with high superconducting transition temperatures
T$_{\text c}$ exhibit abnormal electrical transport
and optical properties. The observed linear temperature
variation of the resistivity was attributed by Lee and Read \cite{lee} to
electron-electron collisions
on a perfectly nested Fermi surface in the form of a square, and the anomalous
linear frequency variation of the damping
has been derived for a partly nested Fermi liquid (NFL).\cite{viro,ruv}
However, nesting models need to consider the
competing spin density wave (SDW) instability.

		The evident correlation of high T$_{\text c}$ values in
cuprates with departures from the standard Fermi liquid behavior
suggests that the physical origin of the anomalous damping is a key source
of the superconductivity. This connection is found here
by the present microscopic
theory for a partially nested Fermi surface.
The on-site Coulomb repulsion $U$
provides the primary interaction, while the nested
orbit topology is the key determinant of the attraction.
The pairing symmetry and binding via exchange of spin fluctuations
are determined by the nesting vector $\vec{Q}$. The d angular
momentum state is found to be favored in the Bi2212 cuprate where our
calculations reveal the existence of
superconductivity for values of U that avoid the SDW instability.

		Our investigation was inspired by the discovery of Scalapino
{\it et al.} \cite{scal} that d-wave
pairing may be possible in a Hubbard model. Similar correlations were found in
Monte Carlo simulations \cite{hirs} on a small lattice. However, for the band
fillings and Fermi
surface topologies treated by Scalapino \cite{scal} and others
\cite{kato,shima,hanke},
the lowest order estimates give very small
T$_{\text c}$ values. Other anisotropic pairing proposals
have been applied to organic compounds \cite{emery} and heavy
Fermion superconductors \cite{ott}.

We consider the Hubbard Hamiltonian
\beq
H= \sum_{\vec{k},\sigma} E(\vec{k})c_{\vec{k},\sigma}^{\dagger}c_{\vec{k},
\sigma}^{} +
U \sum_{\vec{p},\vec{q},\vec{k}} c_{\vec{p}+\vec{q},\uparrow}^{\dagger}
c_{\vec{p},\uparrow}^{}
c_{\vec{k}-\vec{q},\downarrow}^{\dagger}c_{\vec{k},\downarrow}^{},
\label{hamil}
\eeq
where the electron (or hole) energy band $E(k)$ is represented by the
tight--binding expression
\beq
E(\vec{k}) = -2t ( \cos k_x + \cos k_y - B \cos k_x\cos k_y +
\frac{\mu}{2}),
\label{band}
\eeq
$U$ is the Coulomb repulsion between electrons at a given site, and
$c_{\vec{k},\sigma}^{\dagger}
(c_{\vec{k},\sigma})$ represent creation (destruction) operators of
momentum $k$ and spin $\sigma$.
The Fermi surfaces for this model are shown in Figure \ref{fermi} for
a rounded orbit (Fermi liquid FL), and a
nested surface that resembles photoemission experimental data
by Dessau {\it et al.} \cite{dess}
and Shen {\it et al.} \cite{shen} on Bi$_2$Sr$_2$CaCu$_2$O$_8$.

Electron scattering in the singlet spin channel involves a direct Coulomb
term and the
exchange of spin fluctuations \cite{berk} of the form shown in Figure 2.
The phase space for the scattering and
the susceptibility enhancement near the nesting vector are
important for the d-wave pairing as well as for the
SDW instability condition $U\chi'=1$.

    We compute the real part of the susceptibility, $\chi'(\vec{q},\omega)$,
using the standard definition \cite{agd} and include self energy
corrections of the NFL form \cite{viro}. The results for the $E(k)$ model
chosen to represent Bi2212
are shown as a function of momentum in Figure 3.
The double peak structure at low
frequencies is similar to the neutron spectra for the
La$_{2-x}$Sr$_{x}$CuO$_4$ superconductor \cite{aeppli}.
Another
consequence of nesting is the scaling of the spin susceptibility as a function
of $\omega/T$ which has been confirmed by neutron scattering on several
cuprates
\cite{jrsci}.
The SDW constraint on the
susceptibility
requires $U \leq 1.1$ eV in the case of Bi2212 where the bandwidth
is estimated to be 1.5 eV from photoemission data.

Decomposition of the two-particle scattering into angular momentum channels
yields an effective pairing coupling \cite{scal}
\beq
\lambda_l=-{\sum_{\vec{k}\vec{k}'}  \ g_l(\vec{k}) \ V(\vec{k},\vec{k}') \
g_l(\vec{k}')  \ \delta[E(\vec{k})]
 \ \delta[E(\vec{k}')] \over \sum_{\vec{k}}g^2_l(\vec{k}) \
 \delta[E(\vec{k})]}.
\label{lambda}
\eeq
The conventional symmetry classification of the basis set $g_l$
\cite{scal,hirs,kato,shima,hanke}
is $g_s=1$ for the s-wave states, $g_p=\sin k_x$ for p-waves and
the d-wave states with
$g_{x^2 -y^2}= \cos k_x -\cos k_y$ and $g_{xy}= \sin k_x\sin k_y$.

The primary pairing interaction $V(\vec{k},\vec{k}')$ (see Fig. \ref{diagrams})
is given
by the term
with two spin fluctuation bubbles, $U^3 \chi'^2(\vec{k}-\vec{k}')$,
and the exchange term proportional to $U^2\chi'(\vec{k}+\vec{k}')$.
If $\chi'(\vec{q},0)$ is approximately constant, an inspection of
Eq. \ref{lambda}
reveals that superconductivity by this mechanism
is not possible because $\lambda <0$.

To calculate
the coupling, we represent the susceptibility by a
Gaussian form
\beq
\chi'(\vec{q},0) = A + B(\vec{Q})\exp \left [-\frac{(|q_x|-Q_x)^2 + (|q_y| -
Q_y)^2}{2\alpha^2} \right ]
\label{gauss}
\eeq
where $\vec{q}=\vec{k}-\vec{k'}$, $\vec{Q}=(\pi,\pi)$,
and the constants $A$ and
$B(\vec{Q})$ determine the normalization for the
Gaussian.
This model yields a reasonable fit to the computed Bi2212 susceptibility
shown in Figure \ref{susc}.
The actual nesting peaks for our Bi2212 model are at $\vec{Q}_1= (\xi
\pi,\pi)$,
$\vec{Q}_2=(\pi,\xi \pi)$, $\vec{Q}_3=(2\pi-\xi\pi,\pi)$
and $\vec{Q}_4=(\pi,2\pi-\xi\pi)$, with $\xi = 0.91$, but
this four-peak structure produces only small
corrections to the T$_{\text c}$ determined by the simple
Gaussian in Eq. \ref{gauss}.

Our previous analytic derivation \cite{viro} of the NFL susceptibility
using the nesting approximation $E(\vec{k} +\vec{Q})\cong -E(\vec{k})$
gave a logarithmic temperature variation of $\chi'_{\text{NFL}}(\vec{Q},0)$.
However, the present calculation
gives a smaller susceptibility
(see Fig. \ref{susc}), with a weaker temperature dependence.
The susceptibility reduction is caused by the rounded
corners in our Fermi surface model and the influence of
the NFL self-energy $\Gamma=$ Max$(T,|\omega|)$.
Together,
these features avoid the SDW formation at intermediate values
of the interaction, such as $U\simeq 0.96$ eV (compared to the
bandwidth $8t=1.5$ eV) that is used
here.

Numerical integration over momenta gives the coupling $\lambda_{x^2-y^2}$,
and our computed susceptibility indicates an energy cut-off
$\omega_{\text c}\simeq 0.3$ eV.
The leading order evaluation
of the superconducting temperature becomes
\beq
T_{\text c}=\omega_{\text c} \exp \left ( {-1 \over \lambda_{x^2-y^2}} \right
).
\eeq

We first find that the
Fermi Liquid topology (dashed curve in Figure \ref{fermi})
gives $\lambda_{x^2-y^2}=0.016$
which corresponds to a vanishing T$_{\text c}$. This
example is qualitatively similar to other cases studied by several groups
\cite{scal,hirs,kato,shima,hanke}.
The random phase
approximation (RPA) contributions
enhance this coupling near the SDW instability \cite{scal,kato,shima,hanke}.

Nesting topologies increase the attraction in a d-wave channel
as we demonstrate for the Bi2212 cuprate model. Using band parameters that
yield the nested Fermi surface in Fig. \ref{fermi}, $U=1.0$ eV
and $\chi'_{\text{max}}=0.92$ eV$^{-1}$,
we obtain a d-wave coupling in the lowest
order $\lambda_{x^2-y^2} = 0.27$
that gives T$_{\text c}$ = 90 K.

If the
Coulomb coupling is of intermediate strength, e.g. $U\chi'= 0.9$, the RPA
enhancement of
the nested case would elevate the coupling to $\lambda_{x^2-y^2} = 2.7$ and
thereby
predict an enormous T$_{\text c}$.
This situation should stimulate further research on vertex corrections and
self energy effects that may offset the RPA series enhancement.

The sensitivity of the coupling to the magnitude of
the nesting vector is illustrated in Figure \ref{tccurve}.
The Bi2201 cuprate
exhibits a low T$_{\text c}$ =  6 K despite having abnormal resistivities and
optical properties in
league with the high T$_{\text c}$ cuprates.
Shen {\em et al.} \cite{shen2} have discovered by photoemission
spectroscopy that this cuprate possesses a
nesting vector close to $0.8 \vec{Q}$, as compared to the ideal half filled
case of $\vec{Q} = (\pi ,\pi )$ and the
Bi2212 situation with a nesting vector of $0.9 \vec{Q}$. This correlation
is compatible with
a sharp drop in the calculated T$_{\text c}$
values as
seen in Figure \ref{tccurve}. Our model
predicts that the spin susceptibility peaks seen in Figure \ref{susc} for the
Bi2212 case should spread apart in Bi2201
and this feature may be tested by neutron scattering measurements.

We do not find superconductivity of $xy$ symmetry for our Fermi surface
geometry.
The $x^2-y^2$ state for the present nesting model
is consistent with photoemission measurements
of the energy gap anisotropy in Bi2212 \cite{shen}.
If the Fermi surface is rotated in other cuprates, as suggested by
photoemission spectra of YBa$_2$Cu$_3$O$_{7-\delta}$ by Liu {\it et al.}
\cite{liu}, then states of other symmetry should be examined in more detail.
We find a vanishing T$_{\text c}$ for
p-wave symmetry pairing using $g_x = \sin k_x$ in both
the Fermi liquid model
and the nested Fermi surface.

Our model for Bi2212 yields a Van Hove singularity in the density
of states that is located 0.04 eV$ \simeq 500$ K above the Fermi energy.
An arbitrary increase of the chemical potential towards the logarithmic
singularity would result in a SDW phase as shown in Fig. \ref{tccurve}

Impurity scattering should be detrimental to anisotropic pairing as well as
for the
SDW formation. By analogy with the Abrikosov-Gorkov theory, \cite{abri}
d-wave suppression
by disorder constrains T$_{\text c}$ in the cuprates \cite{mill}. Similarly,
non-magnetic impurities also impede
the competing SDW transition \cite{zitt}. The case of chromium reveals a
further sensitivity of
the SDW to impurity induced shifts of the band structure \cite{fish}.
Impurities at sites in the copper oxide planes should
be more destructive for d--wave superconductivity than those at
interplanar sites.

The origin of nesting features in cuprates is evident in band structure
calculations \cite{pick} because of the
nearly half-filled d-bands in two dimensions. Logically, the relative
persistence of parallel segments in a given band subjected to
doping may be stabilized by a second band that acts as a charge reservoir.

Theoretical extensions of the present work may be relevant to higher order spin
fluctuation graphs, including the ``spin bag'' variety\cite{kampf}, and the
self-energy and vertex corrections.
Nesting of
a two-dimensional electronic structure produces \cite{viro,ruv} a linear
frequency variation of the quasiparticle damping that
bears similarities to the Luttinger theory \cite{lutt} for a one
dimensional electron gas, which also exhibits remarkable
charge and spin
dynamics \cite{soly}, \cite{ander}.

Nevertheless, nesting
in two dimensions is distinguished by a crossover temperature $T^*$ below
which the electronic response reverts to standard Fermi liquid behavior.
Accordingly, the
concept of a well-defined Fermi surface is valid in the NFL approach,
despite the unusual damping
features that arise above $T^*$ and a corresponding frequency crossover
$\omega^*$
that are determined
by the nesting geometry.

Our analysis may provide a guide to the design of new superconducting
materials.
The primary
ingredients for the d-wave pairing are a Fermi surface topology
with a nesting vector restricted to a narrow range, a
Coulomb repulsion $U$ of intermediate strength, and a planar electronic
structure that accepts
intercalant atoms
with weak impurity scattering of electrons (or holes) in the conducting layers.

We have benefited from discussions with J.P. Collman, D.S. Dessau, D. Huse,
R.B. Laughlin, W.A.
Little, and
Z.X. Shen. We (J.R. and C.R.) appreciate the hospitality of the physics
department at
Stanford University during a visit sponsored by a Sesquicentennial Associate
award from the University of Virginia.

\begin{figure}
\caption{A nested Fermi surface (NFL) shown by the solid curve was calculated
to fit the photoemission data points of Dessau {\em et al.} [16] using the
tight--binding model of Eq. 2 with $B=0.165$ and $\mu=-0.56$. The
nesting vector is $\vec{Q}^* \simeq 0.91(\pi,\pi)$ in this case.
By contrast, the dashed curve for the same value of $B$ but a chemical
potential $\mu=-1.6$ shows a rounded orbit
reminiscent of a standard Fermi Liquid (FL).}
\label{fermi}
\end{figure}

\begin{figure}
\caption{Diagrams for the electron-electron scattering in the singlet channel
show the direct Coulomb repulsion
by a dotted line and the spin fluctuation exchange processes with a bubble
representing the susceptibility. In the d-wave channel for a nested
Fermi surface, the leading order attractive
contributions from the graphs involving the susceptibility are of the
same order,
whereas the direct bare Coulomb repulsion gives no contribution in the
Hubbard model, because $U$
is assumed to be momentum--independent.}
\label{diagrams}
\end{figure}

\begin{figure}
\caption{The calculated real part of the susceptibility for the Bi2212
band parameters
$B=0.165$ and $\mu=-0.56$ is shown as a function of momentum $|\vec{q}|$
along the direction
$q_x=q_y$ by the solid curve for the damping $\Gamma_{\text{NFL}}=$
Max$(T,|\omega|)$, and by
the dashed curve for a damping $\Gamma=0$.
The calculated maximum $\chi'(\vec{Q}^*)\simeq 0.97$ eV$^{-1}$
constrains $U < 1$ eV which
compares with the bandwidth $8t=1.5$ eV that we estimated from the
photoemission data
of Ref. 16.
The dot-dashed curve represents the Gaussian model of Eq. 4 with $\alpha=1.2$.}
\label{susc}
\end{figure}

\begin{figure}
\caption{T$_{\text c}$ values
as a function of the nesting vector $\vec{Q}^*$ calculated
using the tight-binding model of Eq. 2 are shown by the solid curve.
The Bi2212 parameters are $B=0.165$ and $\mu=-0.56$.
The points are the experimental values.
A surface with a nesting vector $\vec{Q}^*\simeq 0.8(\pi,\pi)$
appropriate to the Bi2201 photoemission data [22] was simulated
using $B=0.33$ and $\mu=-1.36$ which lowers T$_{\text c}$.
Intermediate $\vec{Q}^*$ cases were found by linear interpolation
of the band structure. The shaded region designates the spin-density wave
(SDW) regime.}

\label{tccurve}
\end{figure}


\begin{references}
\bibitem{bcs}  J.Bardeen, L.N.Cooper, and J.R.Schrieffer, Phys. Rev. {\bf 108},
 243 (1957).
\bibitem{pwa}  P.W.Anderson and P.Morel, Phys. Rev. {\bf 123}, 1911 (1961).
\bibitem{don} S.Doniach and S.Engelsberg, Phys. Rev. Lett. {\bf 17}, 750
(1966).
\bibitem{berk}
N.F.Berk and J.R.Schrieffer, Phys. Rev. Lett. {\bf 17}, 433 (1966).
\bibitem{legg}  A.J.Leggett, Rev.
Mod. Phys. {\bf 47}, 331 (1975).
\bibitem{lee}  P.A.Lee and N.Read, Phys. Rev. Lett. {\bf 58}, 2692 (1987).
\bibitem{viro}
A.Virosztek and J.Ruvalds, Phys. Rev. B{\bf 42}, 4064 (1990).
\bibitem{ruv}  J.Ruvalds and
A.Virosztek, Phys. Rev. B{\bf 43}, 5498 (1991).
\bibitem{scal}  D.J.Scalapino, E.Loh,Jr., and J.E.Hirsch, Phys. Rev. B{\bf 34},
8190 (1986); {\it ibid} {\bf
35}, 6694 (1987); H.Q.Lin,
J.E.Hirsch, and D.J.Scalapino, Phys. Rev. B {\bf 37}, 7359 (1988).
\bibitem{hirs} J.E.Hirsch, Phys. Rev. Lett. {\bf 54}, 1317 (1985).
\bibitem{kato} M. Kato and K.Machida, Phys. Rev. B{\bf 37}, 1510 (1988).
\bibitem{shima} H.Shimahara and S.Takada,J. Phys. Soc. Japan {\bf 57},
1044 (1988).
\bibitem{hanke} R.Putz, B.Ehlers, L.Lilly, A.Muramatsu and W.Hanke,
Phys. Rev. B {\bf 41}, 853(1990).
\bibitem{emery} V.J.Emery, Synth. Met. {\bf 13}, 21 (1986).
\bibitem{ott} H.R.Ott, H.Rudiger, T.M.Rice, K.Ueda, and J.L.Smith,
Phys. Rev. Lett. {\bf 52}, 1915 (1984);
K.Miyake, S.Schmitt-Rink, and C.M.Varma,
Phys. Rev. B{\bf 34}, 6554 (1986).
\bibitem{dess} D.S.Dessau {\em et al.}, Phys. Rev. Lett. {\bf 71}, 2781 (1993).
\bibitem{shen} Z.X.Shen {\em et al.}, Phys. Rev. Lett. {\bf 70}, 1553 (1993).
\bibitem{agd} A.A.Abrikosov and L.P.Gorkov and I.E.Dzyaloshinski, {\it Methods
of Quantum Field Theory in Statistical Physics} (Prentice-Hall,
New Jersey, 1963).
The susceptibility is given on p. 179.

\bibitem{aeppli} T.E.Mason {\em et al.}, Phys. Rev. Lett. {\bf 71},
919 (1993) and
references cited therein.
\bibitem{jrsci} J.Ruvalds {\em et al.}, Science {\bf 256},
1664 (1992) and references cited therein.
\bibitem{shen2} Z.X.Shen (unpublished).
\bibitem{liu} Rong Liu {\it et al.}, Phys. Rev. B{\bf 46}, 11056 (1992).
\bibitem{abri} A.A.Abrikosov and L.P.Gorkov, Zh. Eksp. Teor. Fiz. {\bf 39},
1781 (1960):
(Sov.Phys.-JETP {\bf 12}, 1243 (1961)).
\bibitem{mill} A.Millis {\em et al.}, Phys. Rev. B{\bf 37}, 4975 (1988).
\bibitem{zitt} J.Zittartz, Phys. Rev. {\bf 164}, 575 (1967).
\bibitem{fish} R.S.Fishman and S.H.Liu, Phys. Rev. B {\bf 45}, 12306 (1992).
\bibitem{pick} W.E.Pickett, Rev. Mod. Phys. {\bf 61}, 433 (1989)
and references cited therein.
\bibitem{kampf}
A.Kampf and J.R.Schrieffer, Phys. Rev. B{\bf 41}, 6399 (1990).
\bibitem{lutt} J.M.Luttinger, Phys. Rev. {\bf 121}, 942 (1961).
\bibitem{soly} J.Solyom, Adv. Phys. {\bf 28}, 201 (1979).
\bibitem{ander} P.W.Anderson, Phys. Rev. Lett. {\bf 67}, 2092 (1991);
Science {\bf 235}, 1196 (1987).

\end{references}
\end{document}